\documentstyle[twocolumn,prl,aps,epsfig,floats]{revtex}
\begin{document}
\draft

\twocolumn[\hsize\textwidth\columnwidth\hsize\csname @twocolumnfalse\endcsname

\title{
Small Energy Scale for Mixed-Valent Uranium Materials
}

\author{
Mikito Koga and Daniel L. Cox
}

\address{
Department of Physics, University of California Davis, CA 95616
}
\date{\today}
\maketitle

\begin{abstract}
We investigate a two-channel Anderson impurity model with a $5f^1$
magnetic and a $5f^2$ quadrupolar ground doublet, and a $5f^2$ excited
triplet.
Using the numerical renormalization group method, we find a crossover
to a non-Fermi liquid state below a temperature $T^*$ varying as the
$5f^2$ triplet-doublet splitting to the 7/2 power.
To within numerical accuracy, the non-linear magnetic susceptibility
and the $5f^1$ contribution to the linear susceptibility are given by
universal one-parameter scaling functions.
These results may explain UBe$_{13}$ as mixed valent with a small
crossover scale $T^*$.
\end{abstract}


]
\narrowtext

The possibility of non-Fermi liquid (NFL) behavior in Ce- and U-based alloys
has been discussed intensively since the discovery of the anomalous
temperature dependence of their resistivity, magnetic susceptibility and
specific heat coefficient.
A candidate for the explanation of the NFL state is the multichannel Kondo
model, widely used for the $f$-shell materials \cite{Cox98}.
For the $f$-shell impurities, the intra-atomic interactions such as
Hund's and spin-orbit couplings have to be taken into account in the
presence of the crystalline-electric field (CEF), giving rise to various types
of scattering of conduction electrons \cite{Nozi80,Koga95}.
Experimental investigations of dilute Ce or U alloys suggested that
single-site effects of such a magnetic ion are important for NFL physics.
In the U case, the observation of a logarithmically divergent
specific heat coefficient \cite{Seaman91} strongly supported the
quadrupolar (two-channel) Kondo scenario \cite{Cox87}.
As evidence of single-site U effects, such a logarithmic anomaly was
observed in the dilute U limit of U$_x$Th$_{1-x}$Ru$_2$Si$_2$
\cite{Ami94}, although the resistivity in this metal cannot be
explained by the simple two-channel Kondo model \cite{Suzuki97}.
\par
According to photoemission and inverse photoemission studies on U
compounds\cite{Allen96}, well-separated atomic-like peaks cannot be seen
in $5f$-spectra, while for Ce compounds the peaks can be described by
an Anderson impurity model.
This implies that $5f$-properties look much more like mixed valence.
Recently, the two-channel Anderson impurity model in the mixed-valent
regime was proposed to account for the NFL physics of UBe$_{13}$
\cite{Schil98}.
In this model, a low-lying $\Gamma_3$ quadrupolar
(non-Kramers) doublet in the $5f^2$ configuration and $\Gamma_6$ magnetic
(Kramers) doublet in the $5f^3$ configuration are taken into account and
excited CEF states are neglected.
Both states mix with each other via the hybridization between the
localized $f$-orbital and conduction band.
The model successfully describes the temperature dependence of the
non-linear magnetic susceptibility observed in UBe$_{13}$ \cite{Rami94}
and in U$_{1-x}$Th$_x$Be$_{13}$ \cite{Aliev95}, suggesting that strong
quantum fluctuations should drive the CEF state of U ions to a
mixed-valent state between U$^{+3}$ and U$^{+4}$ \cite{Aliev95}.
However, except for virtual effects like the Van Vleck magnetic
susceptibility, excited $5f^2$ CEF levels were neglected in the model.
This leaves a large question, however:
how can the small energy scale ($\simeq 10$K) apparent in thermodynamic
data for, e.g., UBe$_{13}$ be reconciled with such a mixed valence model
for which the smallest plausible value is large ($\simeq 150$K)?
\par
In this paper, we show that the inclusion of the excited CEF levels
resolves all the above controversies.
The dynamics of the excited CEF states reduces the crossover temperature
to the two-channel Kondo state significantly even in the mixed-valent
regime.
Using Wilson's numerical renormalization Group (NRG) method
\cite{Wilson75,Krish80} for a cubic CEF case, we show clearly that
a small energy scale $T^*$ is related to the competition between
different types of NFL fixed points, which is very important to explain
the observed non-linear susceptibility within the two-channel Kondo scenario.
The crossover in the Kondo effect is visualized by flows of NRG energy
levels and $T^*$ can be estimated by observing the recovery of the
symmetry associated with the two-channel Kondo model.
We find that $T^*$ obeys a power-law with respect to the CEF splitting
and can be much smaller as the splitting decreases.
In addition, the non-linear magnetic susceptibility is dominated by
the magnetic moment of the excited configuration, and obeys a
one-parameter scaling law set by $T^*$.
We expect this result for the small energy scale to be generic for
ions such as Pr, Tm and U which fluctuate between configurations with
internal degrees of freedom;
the exponent of the power law will vary with the details of the
CEF spectrum of the ground configuration.
\par
In the present work, we study the two-channel Anderson impurity
model in a cubic system, taking into account the first excited
$\Gamma_4$ triplet state explicitly as well as the lowest-lying
$\Gamma_3$ quadrupolar doublet in $5f^2$:
both states are coupled to the $\Gamma_7$ magnetic doublet (the lowest
possible state in $5f^1$), via the hybridization $V$ with the
$\Gamma_8$ conduction electrons.
Note that the choice of $5f^1$ in our work is motivated by
calculational simplicity;
there is no qualitative difference with respect to the
$5f^2$-$5f^3$ picture used in Ref.~\cite{Schil98}.
The corresponding Hamiltonian is given by
\begin{eqnarray}
H & = & \sum_{km} \varepsilon_k c_{km}^{\dag} c_{km}
+ \sum_M E_M |M \rangle \langle M|
\nonumber \\
& + & V \sum_{km \alpha M} \big(C_{\alpha,m;M}
c_{km}^{\dag} |\alpha \rangle \langle M| + {\rm h.c.} \big),
\label{eqn:1}
\end{eqnarray}
where a single-$f$-electron state $|\alpha \rangle$ represents the
$\Gamma_7$ states;
a two-$f$-electron state $|M \rangle$ represents the $\Gamma_3$ and
$\Gamma_4$ states.
The operator $c_{km}^{\dag}$ ($c_{km}$) creates (annihilates) a
$\Gamma_8$ conduction
electron with a wave number $k$, the kinetic energy of which is
given by $\varepsilon_k$.
The energy of the $f$-electron state $E_M$
($= E_{\Gamma_3},E_{\Gamma_4}$) is measured from that of the
$\Gamma_7$ magnetic doublet.
The Clebsch-Gordan coefficient $C_{\alpha,m;M}$ is given by
$\langle \alpha |f_m|M \rangle$, where $f_m$ annihilates an
$f$-electron with $\Gamma_8$ symmetry.
If the energy of the $\Gamma_4$ triplet state is so large that the
contribution from the triplet can be neglected, the Hamiltonian
(\ref{eqn:1}) becomes simpler as presented in Ref.~\cite{Schil98}.
Assuming further that $\pi \rho V^2 \ll |E_{\Gamma_3}|$ ($\rho$ is the
conduction-electron density of states at the Fermi level), the Hamiltonian
is reduced to the two-channel Kondo Hamiltonian with $S = 1/2$
($S$ is the size of a local spin) \cite{Nozi80,Cox87}, which has been
solved by various methods \cite{Cox98}.
We can label the four $\Gamma_8$ states by a tensor product of two
quadrupolar and two magnetic indices.
Each of them has SU(2) symmetry in the restricted Hilbert space of
this isotropic Hamiltonian.
In our model, on the other hand, we have to treat the cubic symmetry
explicitly because mixing of $\Gamma_3$ and $\Gamma_4$ states
prevents the SU(2) labels from being good quantum numbers.
\par
In spite of such a complicated atomic structure of the impurity,
the Hamiltonian (\ref{eqn:1}) can be solved by the NRG method
\cite{Wilson75,Krish80}.
For the purpose of numerical calculation, it is transformed to the
following hopping Hamiltonian with the recursion relation
\begin{eqnarray}
& & H_{N+1} = \Lambda^{1/2} H_N
\mbox{} + \sum_m (s_{N+1,m}^{\dag} s_{Nm} + {\rm h.c.}),
\nonumber \\
& & H_0 = \Lambda^{-1/2} \Big[\sum_M
\tilde{E}_M |M \rangle \langle M|
\nonumber \\
& & ~~~~
\mbox{} + \tilde{\Gamma}^{1/2} \sum_{m \alpha M}
\big(C_{m,\alpha;M} s_{0m}^{\dag} |\alpha \rangle \langle M|
\mbox{} + {\rm h.c.} \big) \Big].
\end{eqnarray}
Here $H_0$ describes the impurity site, and $s_{nm}^{\dag}$ creates
a $\Gamma_8$ electron in the conduction band discretized logarithmically
as $D / \Lambda$ (D is a half width of the band).
The energy of two $f$-electron states $\tilde{E}_M$ and the
hybridization $\tilde{\Gamma}$ are equal to $E_M/D$ and $\rho V^2/D$,
respectively, except for some $\Lambda$-dependent scaling factors
\cite{Krish80}.
We diagonalize the NRG Hamiltonian $H_N$ at each NRG step to solve the
eigenenergies, and we construct $H_{N+1}$ from $H_N$, using the
Clebsch-Gordan method for the cubic symmetry \cite{Koster63},  which
speeds up the calculation ten-fold.
Introducing an axial magnetic field $h_{z,0}$ on the impurity site, we
can calculate the temperature $T$ dependence of the impurity
magnetization by using
\begin{eqnarray}
& & M_z = {{\rm Tr} J_z \exp \left[- \bar{\beta} (H_N + H'_N) \right]
\over {\rm Tr} \exp \left[- \bar{\beta} (H_N + H_N') \right]}.
\end{eqnarray}
Here the axial component $J_z$ of the total angular momentum is coupled
to the magnetic field in the Zeeman term
$H'_N = - g_J J_z h_z$ ($g_J$ is a Land{\'e}'s g factor and we take
$\mu_{\rm B} = k_{\rm B} = 1$).
The physical temperature $T$ and the rescaled magnetic field $h_z$ are
given by
\begin{eqnarray}
& & T = {1 + \Lambda^{-1} \over 2}
\Lambda^{-(N-1)/2} / \bar{\beta},
\nonumber \\
& & h_z = {2 \over 1 + \Lambda^{-1}} \Lambda^{(N-1)/2} h_{z,0},
\end{eqnarray}
respectively, where we take $\bar{\beta} \sim 2$.
The linear susceptibility $\chi_{\rm imp}^{(1)}$ and non-linear
susceptibility $\chi_{\rm imp}^{(3)}$ are calculated by expanding
$M_z$ in $h_{z,0}$ ($M = \chi^{(1)} h + \chi^{(3)} h^3/6 + \cdots$).
In the NRG calculation, these are explicitly given by
\begin{eqnarray}
& & T \chi_{\rm imp}^{(1)} = \bar{\beta}^{-1}
\lim_{h_z \rightarrow 0} \left[M_z (h_z) - M(0) \right] / h_z,
\nonumber \\
& & T^3 \chi_{\rm imp}^{(3)} = 6 \bar{\beta}^{-3}
\lim_{h_z \rightarrow 0}
\left[M_z(h_z) - \bar{\beta} T \chi_{\rm imp}^{(1)} h_z \right] / h_z^3.
\end{eqnarray}
This calculation of the impurity susceptibility gives sufficient
accuracy over a wide range of temperatures, as long as the
hybridization $V$ is restricted to $\pi \rho V^2 \ll D$.
Throughout the NRG calculations, we take $\Lambda = 3$ and keep $500$
or $800$ lowest-lying states at each NRG step.
Although the latter number reduces the numerical error relative to the
former, a qualitative difference does not appear.
We show the former results in this paper.
\par
First we shortly discuss the case $E_{\Gamma_4} \gg E_{\Gamma_3}$ in
the mixed-valent regime \cite{Schil98}.
As the temperature decreases, two characteristic temperatures are obtained:
at a higher temperature, the highest of the $\Gamma_3$ quadrupolar and
$\Gamma_7$ magnetic doublets is screened by the conduction electrons;
at a lower temperature, the other remaining doublet is screened.
The NFL behavior due to the two-channel Kondo effect appears at the
lower temperature $T_{\rm K}$.
$T_{\rm K}$ increases with the increase of $V$ and the decrease of
$|E_{\Gamma_3}|$.
According to the analysis for the realistic parameter region of the
Hamiltonian, the evaluated $T_{\rm K}$ is larger by an order of
magnitude than the 10K observed in UBe$_{13}$.
If the contribution from the $\Gamma_4$ triplet to the Kondo effect
is taken into account, $T_{\rm K}$ is enhanced as the CEF splitting
$\Delta = E_{\Gamma_4} - E_{\Gamma_3} (>0)$ approaches zero,
and a smaller temperature $T^*$ appears at which we find the
crossover from the CEF states to the NFL states.
Below we discuss this energy scale, restricting ourselves to the
realistic case where $f^2$ is the most stable configuration
($E_{\Gamma_3} < E_{\Gamma_4} < 0$) and the two-channel Kondo effect occurs at
low temperatures.
\par
\begin{figure}
\begin{center}
\psfig{figure=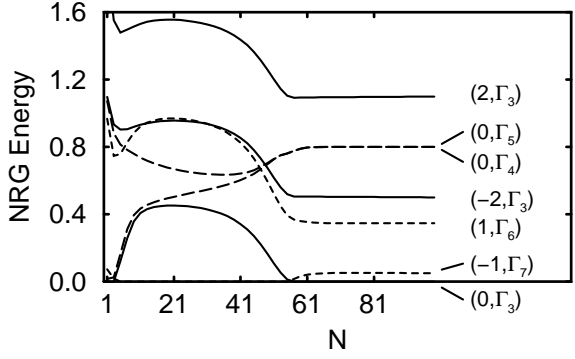,width=7.5cm}
\end{center}
\vspace*{0.05cm}
\caption{
Flow of the most relevant NRG energy levels for the odd number of NRG
steps $N$ and $\Lambda =3$.
We take $\tilde{\Gamma} = 0.15$, $\tilde{E}_{\Gamma_3} = -0.200$ and
$\tilde{E}_{\Gamma_4} = -0.168$ here.
Each energy level is labeled by $(Q,\Gamma_i)$, where $Q$ represents
the total number of particles measured with respect to the ground state.
At the fixed point, the $(-1,\Gamma_7)$ and $(1,\Gamma_6)$ doublets
have a center of gravity energy of $0.2$, while the degenerate
$(0,\Gamma_4)$ and $(0,\Gamma_5)$ triplets have an energy of 0.8.
These values correspond to the first and second excited NRG energies
given by the two-channel Kondo model for $\Lambda = 3$
\protect\cite{Pang91}.
This implies that only particle-hole symmetry breaking causes the
splitting of the $(-1,\Gamma_7)$ and $(1,\Gamma_6)$ doublets and that of
the $(-2,\Gamma_3)$ and $(2,\Gamma_3)$ doublets.
}
\end{figure}
Our NRG results show that the stable fixed points are classified into
two types:
for the first, the low-lying excitations can be described by
assuming a $\Gamma_3$ local moment coupled to the conduction electrons.
For the second, a $\Gamma_4$ local moment is dominant in the
low-lying excitations.
The stability of the fixed points depends on whether the bare CEF splitting
$\Delta$ is above or below a critical value $\Delta_{\rm c}$ ($> 0$).
The following discussion is restricted to the case
$\Delta' = \Delta - \Delta_{\rm c} >0$ where we obtain the
first fixed point, described by the two-channel Kondo model,
as is easily found from a flow diagram of the low-lying NRG energy
levels in Fig.~1.
For a small number of NRG steps, the almost degenerate
(0,$\Gamma_3$) and (0,$\Gamma_4$) states correspond to the
lowest-lying CEF states with small splitting $\Delta$.
As the number of NRG steps increases, the splitting of (0,$\Gamma_3$)
and (0,$\Gamma_4$) levels becomes larger, and the former becomes
the lowest state at the fixed point (where all the NRG energy levels
reach constant values).
The double degeneracy of $\Gamma_3$ corresponds to that of a local
spin of the two-channel Kondo model \cite{Cox87}.
Practically we define $T^*$ as the temperature corresponding to
the number of NRG steps at which the difference of the NRG energy levels
between the (0,$\Gamma_4$) and (0,$\Gamma_5$) triplets reduces to $0.01$.
The degeneracy of these states is one piece of evidence for the
two-channel Kondo fixed point.
\par
\begin{figure}
\begin{center}
\psfig{figure=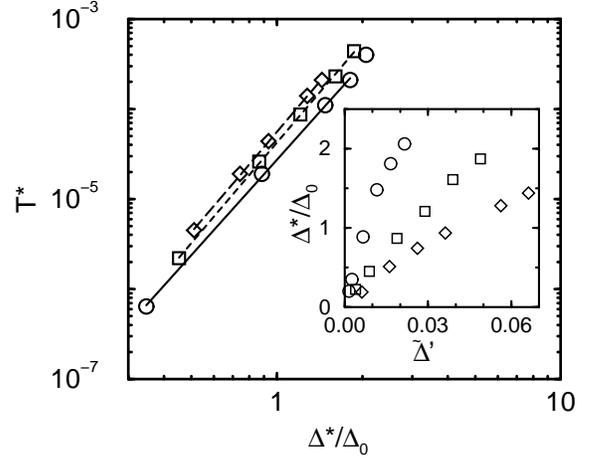,width=7.5cm}
\end{center}
\vspace*{0.05cm}
\caption{
Relation between the small energy scale $T^*$ and the effective
CEF splitting $\Delta^*$ for $\tilde{\Gamma} = 0.10$ ($\circ$), $0.15$
($\Box$) and $0.20$ ($\Diamond$), where we fix $\tilde{E}_{\Gamma_3}$
at $-0.200$.
The corresponding lines are represented by
$T^* = a(\Delta^* / \Delta_0)^b$, given as
$(a,b) = (2.7 \times 10^{-5},3.5)$, $(4.2 \times 10^{-5},3.7)$
and $(5.6 \times 10^{-5},3.7)$, respectively.
Inset: $\Delta^* / \Delta_0$ versus $\tilde{\Delta}'$
($= 3[\Delta - \Delta_{\rm c}] / 2$).
For the above ordering of $\tilde{\Gamma}$ values, $3\Delta_{\rm c}/2$
is 0.018, 0.031 and 0.044.
}
\end{figure}
Figure~2 shows that $T^*$ is proportional to the power law with an effective
CEF splitting $\Delta^*$.
In analogy with Boltzmann weight factors, $\Delta^*$ is associated with
the occupancy weights of $f$-electrons $n_3^*$ and $n_4^*$ in the
$\Gamma_3$ doublet and $\Gamma_4$ triplet at the fixed point, respectively:
\begin{eqnarray}
& & \Delta^* = \Delta_0 \ln \left({3 n_3^* \over 2 n_4^*}\right).
\end{eqnarray}
Here $\Delta_0$ depends on only the hybridization $\tilde{\Gamma}$
if the other parameters are fixed.
When $\Delta^*$ goes to zero with $\Delta'$, both the $\Gamma_3$
doublet and $\Gamma_4$ triplet become relevant with the same weight.
Each line in Fig.~2 has almost the same power, implying
$T^* \propto (\Delta^*)^{7/2}$.
As $\Delta'$ approaches zero (we use the NRG scale
$\tilde{\Delta}' = 2/[1 + \Lambda^{-1}]\Delta'$ in Fig.~2), it tends to
be proportional to $\Delta^*$ and satisfies
$T^* \propto (\Delta')^{7/2}$.
This relation can be derived from the equation for a small energy scale
$T_{\rm CEF}^{\rm x} \simeq T_0 (\Delta / T_0)^2 /4$ appearing in Sec 5.3.2
of Ref.~\cite{Cox98}.
Since $T_0$ is the temperature at which the $\Gamma_3$ doublet is screened
by the conduction electrons, written as
$T_0 = (D/\Delta)^{3/2} T_{{\rm K},\Gamma_3}$, where
$T_{{\rm K},\Gamma_3}$ is the Kondo temperature for the case without
the higher $\Gamma_4$ triplet \cite{Yamada92}, we obtain
\begin{eqnarray}
& & T_{\rm CEF}^{\rm x} \simeq {1 \over 4} {D^2 \over T_{{\rm K},\Gamma_3}}
\left({\Delta \over D} \right)^{7/2}.
\label{eqn:3}
\end{eqnarray}
This resembles the relation of $T^*$ and $\Delta'$ if $\Delta'$ is
small.
Actually the hybridization renormalizes $\Delta$ to $\Delta'$
as obtained by our NRG calculation.
For the two-channel Kondo effect, $T_{{\rm K},\Gamma_3}$ is given by
$\rho V_{\rm eff}^2 \exp (E_{\Gamma_3} / 2\rho V_{\rm eff}^2)$, where
the ratio $V_{\rm eff}^2 / V^2 = 8/21$ comes from a Clebsch-Gordan
coefficient when we derive the effective exchange interaction due to
the $\Gamma_3$ doublet.
Using Eq.~(\ref{eqn:3}), we can roughly estimate the ratio
$T^*/(\Delta')^{7/2}$ as $1/(10^6 T_{{\rm K},\Gamma_3})$, where $D$ is
taken to be unity.
\par
In Fig.~3, we show the temperature dependent magnetic susceptibility
rescaled by $T^*$.
The linear susceptibility $\chi_{\rm imp}^{(1)}$ is approximately
given as a sum of Zeeman and Van Vleck parts.
The former term is dominated by the $\Gamma_7$ magnetic susceptibility
$\chi_{{\rm imp},\Gamma_7}^{(1)}$.
Each curve $\chi_{{\rm imp},\Gamma_7}^{(1)}$ is well fitted by a
function $f(T/T_{\rm s})/T_{\rm s}$, logarithmic below
$T \simeq T_{\rm s}$ ($T_{\rm s} \simeq 10 T^*$).
On the contrary, such a simple scaling function cannot be found
for the Van Vleck contribution, which is expected to be given by
$a - b \sqrt{T/T_{\rm s}}$ \cite{Cox94}.
In this case, $a$ and $b$ appear to be complicated functions of $T^*$
and $\Delta^*$.
Since $|\langle \Gamma_3 |J_z| \Gamma_4 \rangle|^2 / \Delta$ is very
large in our model, mainly the Van Vleck part contributes to
$\chi_{\rm imp}^{(1)}$, and thus a simple scaling relation for
$\chi_{\rm imp}^{(1)}$ cannot be found.
\par
\begin{figure}
\begin{center}
\psfig{figure=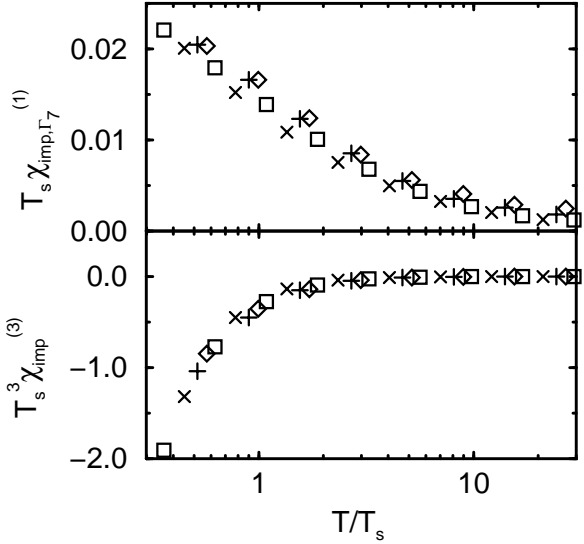,width=7.5cm}
\end{center}
\vspace*{0.05cm}
\caption{
Temperature dependent impurity magnetic susceptibility.
The curves are for $\tilde{\Gamma} = 0.10$, $\tilde{E}_{\Gamma_3} = -0.200$,
and $\tilde{E}_{\Gamma_4} = -0.160$ ($\Diamond$),
$-0.165$ ($+$), $-0.170$ ($\Box$) and $-0.175$ ($\times$), for which
$T_{\rm s}$ ($\simeq 10T^*$) is, respectively, $3.6 \times 10^{-3}$,
$2.3 \times 10^{-3}$, $1.1 \times 10^{-3}$ and $1.7 \times 10^{-4}$,
and the $5f^2$ occupancy in the ground state is $0.775$, $0.765$, $0.751$
and $0.727$, respectively.
Upper curve: The $5f^1$ contribution to the linear susceptibility
$\chi_{\rm imp}^{(1)}$.
Lower curve: The total non-linear susceptibility.
}
\end{figure}
On the other hand, the curves for the non-linear susceptibility
$T_{\rm s}^3 \chi_{\rm imp}^{(3)}$ do scale with $T / T_{\rm s}$, implying
that the main contribution comes from the $\Gamma_7$ magnetic doublet even
when $f^2$ is more stable than $f^1$ ($0.5 < n_3^* + n_4^* \lesssim 0.8$).
Usually the $\Gamma_3$ quadrupolar moment gives rise to a positive
$\log (T_{{\rm K},\Gamma_3}/T)$ dependence in
$\chi_{\rm imp}^{(3)}$ due to the two-channel Kondo effect, which
should be caused by the Van Vleck coupling of the $\Gamma_3$ doublet
and $\Gamma_4$ triplet.
However, the magnetic part derived from the $\Gamma_7$ behaves as
$-1/T^3 \rightarrow - 1/T$ below $T_{\rm s}$ and overcome the quadrupolar
contribution.
This trend of negative $\chi_{\rm imp}^{(3)}$ is in good agreement
with the non-linear susceptibility derived by the simple model
where the dynamics of the $\Gamma_4$ triplet was neglected \cite{Schil98}.
However, for comparison of the scaling behavior, $T_{\rm K}$ must be
replaced by $T_{\rm s}$.
For $E_{\Gamma_3} < 0$, $T_{\rm K}$ is approximately equal to
$T_{{\rm K}, \Gamma_3}$ in Eq.~(\ref{eqn:3}) and is very large in the
strongly mixed-valent regime.
On the other hand, $T^*$ ($\simeq 0.1T_{\rm s}$) can be much smaller when
$\Delta$ is close to $\Delta_{\rm c}$.
We conclude that $T^*$, characteristic to the competition between the
two different types of NFL fixed points, can represent the small energy
scale of around $10$K for UBe$_{13}$.
\par
As discussed above, the low-energy physics below $T^*$ is described
very well by the two-channel Kondo model as long as the CEF splitting
satisfies $\Delta > \Delta_{\rm c}$.
In the case $\Delta < \Delta_{\rm c}$, low-lying excitations dominated by
the relevant triplet at low temperatures are very different
from those given by the two-channel Kondo model.
The study of this case is also required and motivated by a recent
experiment of U$_x$Y$_{1-x}$Pd$_3$ where a $\Gamma_5$ triplet is
almost degenerate with a $\Gamma_3$ doublet \cite{Bull98}.
We anticipate that a small crossover scale will still result for
$\Delta < \Delta_{\rm c}$, albeit with a different power law exponent.
Indeed, this result should be generic for mixed-valent ions such as
Pr and Tm, having configurations with internal degrees of freedom
for the lowest CEF levels.
\par
We are grateful to A. Schiller for stimulating discussion.
K. M. is supported by JSPS Postdoctoral Fellowships for Research
Abroad.
We also acknowledge the support of the US Department of Energy, Office
of Basic Energy Research, Division of Material Science.

\end{document}